\documentclass[conference]{IEEEtran}
\IEEEoverridecommandlockouts
\usepackage{cite}
\usepackage{balance}
\usepackage{changes}
\usepackage{amsmath,amssymb,amsfonts}
\usepackage{algorithmic}
\usepackage{tablefootnote}
\usepackage{graphicx}
\usepackage{multirow}
\usepackage{array}
\usepackage{textcomp}
\usepackage[hidelinks]{hyperref}
\usepackage{url}
\usepackage{changes}
\usepackage{float}
\usepackage[numbers, sort&compress]{natbib}
\usepackage{pifont}

\usepackage{xcolor}
\def\BibTeX{{\rm B\kern-.05em{\sc i\kern-.025em b}\kern-.08em
    T\kern-.1667em\lower.7ex\hbox{E}\kern-.125emX}}

\usepackage{enumitem}
\setlist[itemize]{leftmargin=10pt}
    
\begin{document}

\title{Deep Reinforcement Learning for Intrusion Detection in IoT: A  Survey}

\makeatletter 
\newcommand{\linebreakand}{%
  \end{@IEEEauthorhalign}
  \hfill\mbox{}\par
  \mbox{}\hfill\begin{@IEEEauthorhalign}
}
\makeatother 

\author{\IEEEauthorblockN{Afrah Gueriani}

\IEEEauthorblockA{\textit{LSEA Lab., Faculty of Technology} \\
\textit{University of MEDEA}\\
Medea 26000, Algeria\\
gueriani.afrah@univ-medea.dz }
\and
\IEEEauthorblockN{Hamza Kheddar}
\IEEEauthorblockA{\textit{LSEA Lab., Faculty of Technology} \\
\textit{ University of MEDEA}\\
Medea 26000, Algeria \\
kheddar.hamza@univ-medea.dz}
\and 
\IEEEauthorblockN{Ahmed Cherif Mazari}
\IEEEauthorblockA{\textit{LSEA Lab, Faculty of Science} \\
\textit{ University of MEDEA}\\
Medea 26000, Algeria \\
mazari.ahmedcherif@univ-medea.dz}
}

\makeatletter

\def\ps@headings{%
\def\@oddhead{\parbox[t][\height][t]{\textwidth}{\flushleft

\noindent\makebox[\linewidth]
}
\vspace{0.5cm}
\hfil\hbox{}}%
\def\@oddfoot{\MYfooter}%
\def\@evenfoot{\MYfooter}}

\def\ps@IEEEtitlepagestyle{%
\def\@oddhead{\parbox[t][\height][t]{\textwidth}{\centering
2023 2nd International Conference on Electronics, Energy and Measurement (IC2EM 2023)\\
}\hfil\hbox{}}%
\def\@oddfoot{ 979-8-3503-1424-3/23/\$31.00 \textcopyright 2023 IEEE \hfil 
\leftmark\mbox{}}%
\def\@evenfoot{\MYfooter}}

\maketitle

\begin{abstract}
The rise of new complex attacks scenarios in Internet of things (IoT) environments necessitate more advanced and intelligent
cyber defense techniques such as various Intrusion Detection Systems (IDSs) which are responsible for detecting and mitigating malicious activities in IoT networks without human intervention. To address this issue, deep reinforcement learning (DRL) has been proposed in recent years, to automatically tackle intrusions/attacks. In this paper, a comprehensive survey of DRL-based IDS on IoT is presented. Furthermore, in this survey, the state-of-the-art DRL-based IDS methods have been classified into five categories including wireless sensor network (WSN), deep Q-network (DQN), healthcare, hybrid, and other techniques. In addition, the most crucial performance metrics, namely accuracy, recall, precision, false negative rate (FNR), false positive rate (FPR), and F-measure, are detailed, in order to evaluate the performance of each proposed method. The paper provides a summary of datasets utilized in the studies as well.

\end{abstract}

\begin{IEEEkeywords}
intrusion detection system, deep reinforcement learning, internet of things, wireless sensor network.\end{IEEEkeywords}

\section{Introduction}
\label{sec1}

The Internet of things (IoT) has evolved significantly in the past few years, allowing the interconnection of multiple devices as well as improving communication and data exchange. IoT has become increasingly integral to modern daily life through enhancing connection, efficiency, and comfort in numerous domains. IoT transforms the way we interact with our surroundings, by simply linking devices, sensors, and systems. It has affected users' daily lives in many ways like industrial control systems (ICSs) which are used to connect, monitor, and control physical processes in industrial environments such as manufacturing, smart grid, smart transportation, and connected agriculture. IoT has brought healthcare and tele-medicine opportunities by advancing and perfecting services.
IoT devices are vulnerable to plenty of cyber threats, including intrusion and illegal access, thus this interconnection also brings up severe security and privacy concerns \cite{kheddar2023deep}. IoT environments are threatened by many attacks and malicious activities that might affect the integrity and security of connected devices and networks. In this context, IDSs become particularly crucial because of their importance in mitigating threats. An IDS is software that continually scans the network for activities that might indicate an impending assault \cite{chang2022survey}.
IDSs serve an essential role in protecting IoT networks by detecting and mitigating attacks and responding to threats in real-time \cite{umer2022machine, chang2022survey}. The IDSs are commonly created with AI methods like machine learning (ML) and data mining (DM) approaches \cite{kheddar2023deep}. One such approach of ML is the utilization of reinforcement learning (RL). 
In order to develop efficient IDSs in IoT environments, deep reinforcement learning (DRL), a sub-field of ML, has emerged as a potential method. DRL combines the strengths of DL which enables the model to learn complex representations and patterns and allows the system to make sequential choices. 

DRL has a huge effect on the intrusion detection field, where it is able to adapt and learn from the changing dynamics of IoT networks. DRL-based IDSs also can autonomously make decisions on whether network activities are normal or indicative of intrusion attempts. The benefits of using RL in IDS include the adaptability to evolving attack patterns, real-time decision-making capabilities, and proficiency in handling complex threats. The integration of RL into IDS holds the potential to considerably improve their efficacy, precision, and responsiveness in identifying and countering cyber threats.

Many suggested state-of-the-art (SOTA) techniques of RL/DRL-based IDS algorithms for IoT are discussed in-depth in this survey.

\subsection{Related Work}
Various IoT environments and WSNs have been taken into consideration in  IDS research. In \cite{rana2022intrusion}, a brief overview of IDS in cloud computing (CC) is provided. The authors review the challenges and risks linked to CC environments. They discuss various IDS's approaches and their ability to identify and mitigate intrusions. Moving on, valuable details about IDSs were discussed in \cite{vinolia2023machine}. The paper covered a review of IDS approaches using both ML, and DL in cloud environments where the authors explore multiple IDS techniques in this area. A comprehensive review of rule learning-based IDS and their potential applications in the context of smart grids (SGs) are presented in \cite{liu2021review}. The paper examines multiple elements related to IDS, such as techniques for feature selection, evaluation metrics, and rule development. It underlines how much is important the rule learning-based IDS in identifying and mitigating security threats in SGs.

In this survey, we aim to provide an overview of the research contributions made in the field of IDSs in IoT environments. The paper focuses on analyzing and highlighting the importance of using DRL to create IDS models and discover attacks in order to protect IoT networks. This survey also opens the opportunity to take a look at the latest research and continue the challenges in the field. The papers considered in our study were chosen regarding their quality as academic articles (indexed at least in the Scopus database) in IoT IDS, as well as the year they were published (between 2018 and 2022). Table \ref{tab:related-work} summarizes the comparison of our proposed review against some existing surveys. It is obvious that the proposed review has addressed all the fields, including the application of DRL-based IDS, datasets description, metrics, the taxonomy of IDS, the taxonomy of RL, and future directions, While the majority of other investigations have either disregarded these domains or only partially addressed them.

This survey is structured as follows: Section \ref{sec2} gives preliminaries that contain standard benchmark datasets and several metrics.   Section \ref{sec3} reviews the existing SOTA methods for DRL-based IDS in IoT environments. Sections \ref{sec4}, and \ref{sec5} covered future direction and conclusion, respectively.

\begin{table*}[h!]
\caption{Related work comparison of the proposed survey against other existing IDS reviews and surveys.}
\label{tab1}
\begin{tabular}{llllllll}
\hline
Ref. & Description & IDS datasets & Metrics & Taxonomy of IDS & Taxonomy of RL & DRL-based IDS & Future directions \\
\hline
\cite{rana2022intrusion} & IDS in CC & \ding{51} & \ding{51} & \ding{55} & \ding{55} & \ding{55} &\ding{51} \\[0.6mm]
\cite{vinolia2023machine} & ML and DL-based IDS in CC & \ding{51} & \ding{55} & \ding{55} & \ding{55} & \ding{55} & \ding{51}\\
\cite{liu2021review} & rule learning-based IDS in SG & \ding{55} & \ding{55} & \ding{51} & \ding{55} & \ding{55} & \ding{55}\\

Ours & DRL based IDS in IoT & \ding{51} & \ding{51} & \ding{51} & \ding{51} & \ding{51} & \ding{51} \\

\hline
\end{tabular}
\label{tab:related-work}
\end{table*}

\section{Preliminaries}
\label{sec2}

\subsection{Datasets}

\begin{itemize}
\item \textbf{NSL-KDD \footnote{\url{https://www.unb.ca/cic/datasets/nsl.html}}:} This dataset is the newest version of KDD' 99, the advantages of this dataset over the KDD' 99 are as follows: (1) It excludes redundant records in the train set, (2) The proposed test sets include no duplicate records. The original records of attacks in the KDD train set is  3925650 and the normal record is 972781, while, the original records in the KDD test set is 250436 in terms of attacks and 60591 in terms of normal original records. This dataset contains user-to-root (U2R) and remote-to-local (R2L) attacks.

\item \textbf{UNSW-NB15\footnote{\url{https://research.unsw.edu.au/projects/unsw-nb15-dataset}}:} 
Is created by the IXIA PerfectStorm tool, the tcpdump tool, the Argus, and Bro-IDS tools, in order to produce attacks. Fuzzers, analysis, backdoors, denial of service (DoS), exploits, generic, reconnaissance, shell code, and worms are among the nine attack categories in this dataset. There are two million and 540044 records overall in the UNSW-NB15 dataset. 
\item \textbf{CICIDS-2017\footnote{\url{https://www.unb.ca/cic/datasets/ids-2017.htmlt}}:} 
It implements different types of attacks such as Brute Force FTP, brute force SSH, DoS, heartbleed, web attack, infiltration, botnet, and DDoS. 
\item \textbf{CIC-DDoS2019\footnote{\url{https://www.unb.ca/cic/datasets/ddos-2019.html}}:} 
Various current reflected DDoS attacks are included in this dataset, notably PortMap, NetBIOS, LDAP, MSSQL, UDP, UDP-Lag, SYN, NTP, DNS, and SNMP attacks. 

\item \textbf{BoTNeTIoT-L01\footnote{\url{https://www.kaggle.com/datasets/azalhowaide/iot-dataset-for-intrusion-detection-systems-ids}}:} 
The new version of the data set reduces redundancy by selecting features from a 10-second time window only. The dataset's class label employs a scale of zero for attacks and one for normal samples.

\item \textbf{N-BaIoT\footnote{\url{https://www.kaggle.com/datasets/mkashifn/nbaiot-dataset}}:} 
This dataset addresses the lack of public botnet datasets, especially for the IoT. It has multiple characteristics such as: (1) multivariate and sequential, (2) real number of attributes: 115, (3) associated tasks: classification, clustering, (4) number of instances: 7062606. However, as the malicious data can be divided into 10 attacks carried by 2 botnets, the dataset can also be used for multi-class classification.

\end{itemize}

\subsection{Detection metrics}

The DRL-based IDS researchers' community has utilized numerous metrics, many of which have also been employed in other deep learning applications \cite{kheddarASR2023,kheddar2023deepSteg}. These metrics include: 
\begin{itemize}
    \item \textbf{Accuracy and F1-score:} 
    Is a metric which measures the ability of a system to detect normal traffic and harmful traffic. It is calculated using the Equation \ref{eq1}. 
    \begin{equation}
    \small
        \mathrm{Acc}=\mathrm{\frac{TP+TN}{TP+FP+TN+FN}}
        \label{eq1}
    \end{equation} 

     Where, true positive (TP), refers to an accurate IDS, and true negative (TN) represents normal traffic. False positives (FP) are the detections that interpret normal traffic as attacks. False negative (FN), indicates a failure to reveal an intrusion. A robust F1 score could suggest low FP and low FN predictions since it combines precision and recall. It is expressed in Equation \ref{eq6}:

    \begin{equation}
    \small
        \mathrm{F1-Score}= \mathrm{2 \times \frac{Precision \times Recall}{Precision + Recall}}
        \label{eq6}
    \end{equation}
    
    \item \textbf{Recall(Sensitivity, Detection rate):} The detection rate is defined as a parameter that is used to detect correctly the behavior manners that indicate intrusive actions. It can be expressed  using the following formula \ref{eq2}:
  
        \begin{equation}
        \small
        \mathrm{Rc}=\mathrm{\frac{TP}{TP+FN}}
        \label{eq2}
    \end{equation}
    
    \item \textbf{Precision:} Defined as the percentage of records correctly recognizing dangers compared to the total number of estimated records. It is calculated using the Formula \ref{eq3}: 

     \begin{equation}
     \small
        \mathrm{Pr}=\mathrm{\frac{TP}{TP+FP}}
        \label{eq3}
    \end{equation}

    \item \textbf{False negative rate:} Indicates anomalous activities recognized as typical sensor behaviors. 
    It is given by the Formula \ref{eq4}:

    \begin{equation}
     \small
    \mathrm{FNR}=\mathrm{\frac{FN}{FN+TP}}
        \label{eq4}
    \end{equation}

    \item \textbf{False positive rate:} Is a ratio calculated by dividing the total number of normal records by the number of records that were wrongly rejected. 
    It is specified as follows in the Formula \ref{eq5}:
     \begin{equation}
     \small
        \mathrm{FPR}= \mathrm{\frac{FP}{FP+TN}}
        \label{eq5}
    \end{equation}

\end{itemize}

\subsection{Taxonomy of IDS}
 IDSs are categorized into three main types as represented in Figure \ref{fig01}. \textit{\textbf{Anomaly-based}} analyze network traffic and system behavior to identify changes from normal patterns and alert for malicious activities. \textit{\textbf{Signature-based}} compare network traffic or system activities with a database of known attack signatures triggering alerts upon finding a match. \textit{\textbf{Specification-based}} aims to detect known and unknown attacks, but relies on accurate specifications for optimal effectiveness. 

\begin{figure}[h!]
\centering
\includegraphics[scale=0.33]{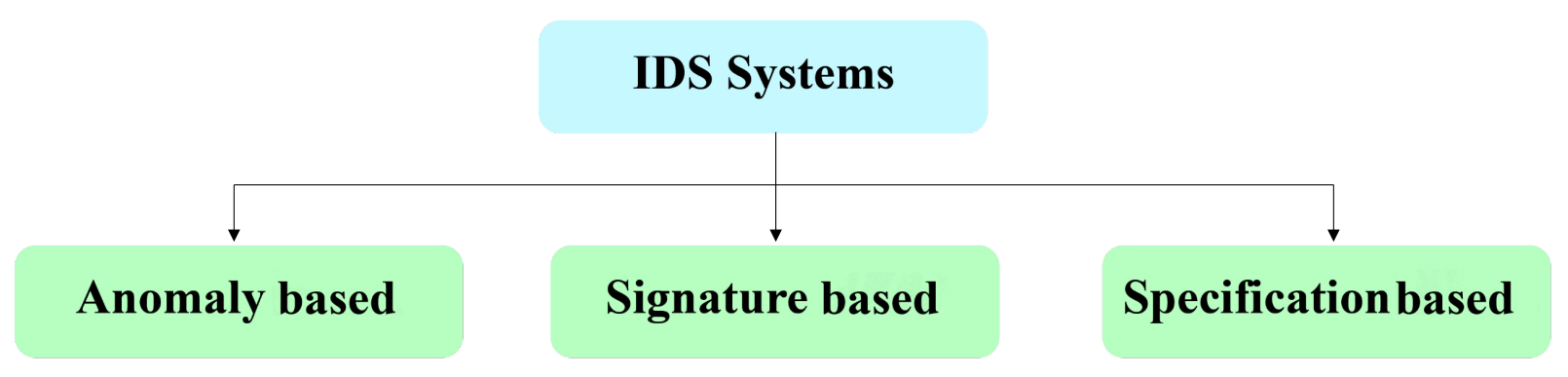}
\caption{Taxonomy of the main categories in IDS.}
\label{fig01}
\end{figure} 

\subsection{Taxonomy of RL}

RL encompasses a wide range of approaches, as shown in Figure \ref{pic07}, which represents a subset of these techniques discussed in our review. \textbf{\textit{Q-learning}} is a model-free RL technique that seeks to determine the best possible reward $(r)$ resulting from taking a specific action $(a)$ in a given state $(s)$. The Q-table acts as the core of RL, summarizing all potential outcomes $r_i(a_i,s_i)$. By exploring various actions and gradually exploiting knowledge, the agent learns to select actions that maximize cumulative rewards within a Markov decision process (MDP) environment. \textbf{\textit{Deep Q-network (DQN)}} is an advancement of Q-learning that replaces the Q-table with a deep neural network (DNN) model to meet real-time requirements. \textbf{\textit{Double deep Q-network (DDQN)}} represents a progressive evolution of DQN, employing a paired DNN architecture. \textbf{\textit{Deep deterministic policy gradient (DDPG)}} is an algorithm that combines DQN and policy gradient (PG) methods, as detailed in \cite{wang2022policy}. This algorithm empowers IDSs to make informed decisions and improve their capability to detect complex and evolving cyber threats effectively. \textbf{\textit{Federated reinforcement learning (FRL)}} is a technique that amalgamates the principles of both RL and federated learning (FL). IDSs utilize this technique to enhance threat detection. Agents learn from distributed data sources, leading to improved accuracy and adaptability in dynamic environments.

\begin{figure}[h!]
\centering
\includegraphics[height=1.5cm, width=9cm]{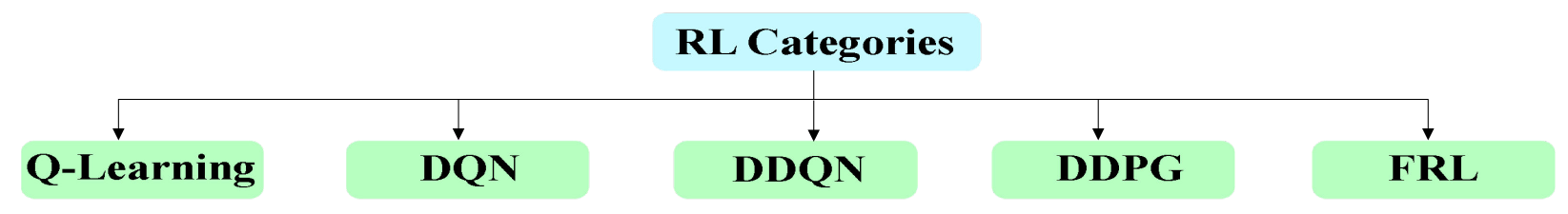}
\caption{Taxonomy of the main categories in RL.}
\label{pic07}
\end{figure}

\section{DRL-based IDS in an IoT environment}
\label{sec3}
DR-based IDS for IoT is divided in the literature into different categories, which are surveyed thoroughly in this section.  Table \ref{tab1} summarizes the performance, the advantages and/or disadvantages of the cited methods for different categories.

\subsection{DRL-based IDS for WSN}

The work in \cite{benaddi2020deep} presents a new approach to enhance security in WSN and IoT by using DRL. The authors address the limitations of traditional IDS by introducing the formalism of MDP to improve performance such as surveying, controlling data networks in real-time, and increasing the quality of surveillance.
The researchers also use the deep Q-network and the NSL-KDD dataset to boost their aims in WSN and IoT. Furthermore, they compared the method against ML k-nearest neighbors (KNN) algorithm. 
In terms of detection accuracy rate and FPR, the experimental evaluation shows that the DRL-IDS surpasses traditional IDS techniques, it works well in detecting various types of attacks such as DoS, DDoS, and intrusion attempts. In \cite{tu2022intrusion} explain an innovative approach to intrusion detection in WSNs that combines generative adversarial networks (GANs) and acro-critic in RL strategies. The authors use GANs and RL algorithms in order to find solutions to the challenging problems of detecting and reducing intrusions and attacks in WSNs. The GAN model produces authentic network traffic data, and an RL agent is used to distinguish between normal and malicious activity. The GAN-generated data is employed to supplement the training dataset, which boosts detection accuracy. 
The findings show that the GAN-RL-based IDS is capable of properly detecting and mitigating intrusions in WSNs. The proposed model was compared with a convolution neural network (CNN), back-propagation (BP), and support vector machine (SVM). The proposed GANs with RL strategies has been found to have great potential in enhancing security measures in WSNs among the other compared methods. The article \cite{diddigi2018novel} discusses the problem of intruder tracking in energy-efficient sensor networks. It provides a unique sensor scheduling approach to reduce the consumption of energy while promising excellent intrusion detection. The authors offer a dynamic sensor scheduling system that intelligently selects a subset of sensors to activate based on the anticipated position of the intruder. The proposed strategy tries to decrease overall network energy usage while retaining adequate coverage to follow the intruder correctly by carefully arranging sensor activations. The framework proposed in \cite{madhurireinforcement} proposes a combined approach that utilizes RL for intrusion detection and cuckoo search technique (CST) for improving the optimal route in WSNs. The cuckoo search algorithm was introduced in this work as an effective optimization technique to improve routing path optimization in WSN. The suggested RLID technique was compared to the BTMA, this technique detects and isolates the intrusion, reducing the latency and has a lower FPR.

\begin{figure}[h!]
\centering
\includegraphics[scale=0.6]{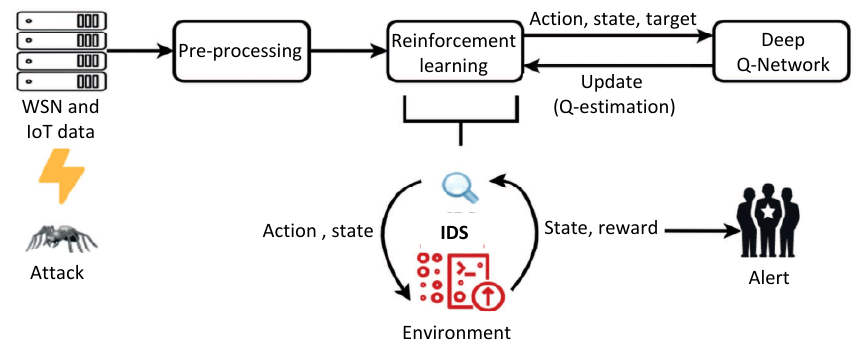}
\caption{Proposed scheme for improving the WSN and IoT based DRL-IDS \cite{benaddi2020deep}.}
\label{fig1}
\end{figure} 

\subsection{Deep Q-Network-based IDS}

 Ramana, M et al. \cite{ramana2022ambient} provide an ambient intelligence approach to improving decision performance in IDSs, when using the IoT. To identify possible attacks, the suggested solution uses the capabilities of IoT devices to collect and examine data from various sources. 
 The authors used DRL with deep Q-neural network in binary attacks and multi-attack classification. In addition, this work contains two steps, the first is to employ the RL approach for initial learning, and the second one is for attack detection and classification.
Metrics like accuracy rate, DR, and false alarm rate (FAR) are used by the authors to assess the decision performance of an IDS. The proposed model was compared with five ML algorithms namely: DQN, MLP, cascade ANN, PCA\_GWO, and GMM-KF, using different datasets.
The proposed approach demonstrates the effectiveness of the ambient intelligence approach in achieving improved real-time decision performance compared to traditional IDSs. Another suggested approach in \cite{nguyen2021federated} proposes deep monitoring, a unique traffic monitoring solution that utilizes FDRL, to address the problem of successfully detecting cybercrimes in IoT networks. To collect data on traffic flow, deep monitor employs software-defined networking (SDN) in IoT networks. The suggested approach dynamically controls flow rules in SDN switches to maximize traffic flow statistic details while avoiding flow table overflow. To enable the deep monitor agents to learn and make decisions in real time depending on the current traffic situation, the authors use DRL algorithms, notably the DDQN.
The FL approach enhances the learning performance of the DDQN algorithm by enabling decentralized collaboration across agents in order to handle the heterogeneity and scalability of IoT systems.
Studies done in an SDN simulation environment show how well deep monitor protects edge devices from the overflow problem and offers fine-grained traffic monitoring capability. Kalnoor et al. in their study \cite{kalnoor2021markov} IDS performance evaluation using a MDP-based model is described in detail, where the use of MDP forms the core of a new framework that accurately captures how an IDS behaves dynamically when it detects and responds to intrusions. Multiple factors are taken into account, including accuracy in identifying threats along with rates of incorrect predictions, when evaluating the efficacy of the IDS. By simulating several attack scenarios, the authors illustrate the utility of their MDP-based approach in measuring the performance of the IDS and adjusting its parameters. The findings offer valuable insights into the design and evaluation of IDS in IoT networks, as well as recommendations for improving their performance. The proposed MDP was compared to CNN and MLP algorithms.

\subsection{Hybrid}

Otoum and Nayak \cite{otoum2021ids} offer a new approach to IDS in the environment of IoT. In this work, the authors deploy a new anomaly and signature
based IDS (AS-IDS) system, which combines anomaly-based and signature-based methodologies to improve IoT network security.
To discover odd behaviors or patterns in IoT materials, the AS-IDS technique applies anomaly detection, which employs deep Q-learning (DQL). The system also includes signature-based detection which compared observed network traffic to known threat signatures.
They thoroughly explain the design and implementation of the AS-IDS system, highlighting its ability to detect five types of attacks in IoT environments. The evaluation of the performance of AS-IDS has been conducted using accuracy and FPR, they also work with NSL-KDD dataset.
The experimental results demonstrate the effectiveness of the AS-IDS system in detecting intrusions and protecting IoT networks. Another hybrid scheme proposed by Tharewal et al. \cite{tharewal2022intrusion} discusses the security concerns posed by the industrial internet of things (IIoT) by employing an IDS based on DRL using proximal policy optimization algorithm (PPO2) and rectified linear unit (ReLU) as an activation function as depicted in Figure 3. This method combines deep learning’s observation capability with DRL, to develop an IDS capable of detecting and reducing intrusions in IIoT networks.
The proposed methodology was compared to ML algorithms such as CNN, LSTM, RNN and to other DRL methods such as DDQN and DQN. The experimental findings and performance assessment data demonstrate how successful the IDS is at detecting intrusions in IIoT networks compared to SOTA methods.

\begin{figure}[h!]
\centering
\includegraphics[width=8cm]{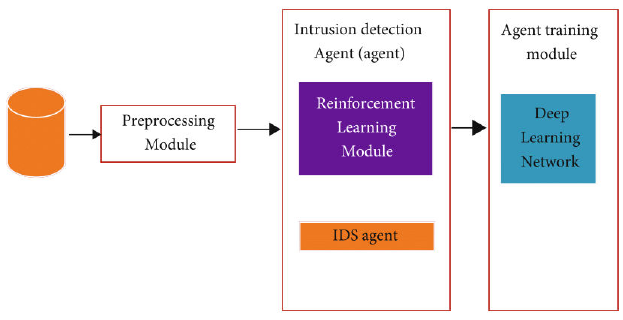}
\caption{Proposed DRL-based IDS for IoT data \cite{tharewal2022intrusion}.}
\label{ppo2}
\end{figure}

\subsection{Application of DRL-based IDS in healthcare} 
Kamble and Gawade \cite{kamble2019digitalization} focus on the digital transformation of healthcare via the use of IoT technology and cryptographic encryption to protect against many types of intrusions, such as DoS and man-in-middle (MIM)  attacks.
The authors determine how IoT devices might be integrated into healthcare systems to improve patient monitoring, data collecting, and communication and offer a scheme to safeguard the transmitted data. Zhang and their colleagues in \cite{otoum2021federated} propose a federated reinforcement learning (FRL)-based IDS to address security challenges in IoT-enabled healthcare systems. Furthermore, the authors raised the existing dangers in IoT devices as well as the necessity of protecting sensitive healthcare data, they introduce an FDRL, where DRL algorithms are used to educate IDS models on distributed IoT devices locally. The models learn and exchange information collectively while protecting data privacy. 
By exploiting the collective intelligence of distributed IoT devices, the recommended IDS improves the system’s capacity to detect and avoid intrusions. Figure \ref{fig4} illustrates the overall architecture of the proposed scheme in \cite{otoum2021federated}.

\begin{figure}[h!]
\centering
\includegraphics[width=8cm]{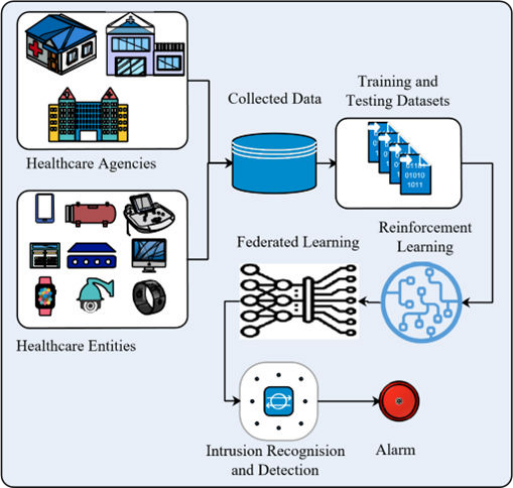}
\caption{An example of proposed DRL-based IDS model for healthcare \cite{otoum2021federated}.}
\label{fig4}
\end{figure}

\begin{table*}[h!]
\caption{List of the surveyed state-of-the-art studies with their advantages and disadvantages.}
\label{tab1}
\begin{tabular}{lllm{1.2cm}m{1.2cm}m{11cm}}
\hline \\
Ref. & Year & Category & Comp. to & Metric& Result / Improvement obtained / Comments / Advantages and/or disadvantages \\
\hline
\cite{benaddi2020deep} & 2020 & WSN & KNN & Accuracy & 98\% / 3.5\%, the proposed model DRL-IDS returns superior performance in terms of detection rate and accuracy with reduced number of false alarms compared to other approaches. The suggested model faces challenges related to computational complexity due to the large number of simulation samples.\\[0.6mm]
\cite{tu2022intrusion} & 2022 & WSN & SVM & Accuracy  & 85.4\% / 19.67\%, comparing the proposed model to three approaches (CNN, BP, and SVM). The combination of GANs with RL method has immense potential for improving security measures in WSNs. Generalization is not demonstrated, as they only use one dataset to assess the effectiveness of the proposed model.\\
\cite{diddigi2018novel} & 2018 & WSN
& ID\_TG & Average track error  & 0.5 / 0.4 
The proposed algorithm combines the advantages of two algorithms ID\_TG and MCTS, with the best performance obtained on a 2D sensor network of size 16 × 16. 
 The paper lacks referencing the necessary metrics, and results-focused only on the average track error.\\

\cite{madhurireinforcement} & 2021 & WSN & BTMA & FNR  & 0.075 / 0.175, the FNR is lower due to the sensor nodes, it is checked by the number of iterations and this provides greater security.\\

\cite{ramana2022ambient} & 2022 & QN & GMM-KF & Accuracy  & 99.4\% / 0.9\%, used numerous datasets and the best results obtained when using IoTID20 dataset.\\

\cite{nguyen2021federated}  & 2021 & QN & FlowStat & DDoS \newline detection  & Compared to the FlowStat solution, the performance of attack detection overall was improved by 22.83.\% \\

\cite{kalnoor2021markov} & 2021 & QN & CNN & Precision  & 99.80\% / 3.05\%, the proposed model RL-IDS gave the
best outcome and precision and AUC leads in multiclass classification. Generalization is not demonstrated, as they only use one dataset to assess the effectiveness of the proposed model, and the comparison is limited to just MLP and CNN techniques.\\
\cite{otoum2021ids} & 2021 & Hybrid & Deep-RNN & DR  & 96.9\% / 15.6\% compared to existing IDS techniques, the suggested AS-IDS model indicates significant improvement. Generalization is not demonstrated, as they only use one dataset to assess the effectiveness of the proposed model in term of accuracy.\\[0.1cm]

\cite{tharewal2022intrusion} & 2022 & Hybrid & DDQN & Accuracy  & 90\% / 9.05\%, the real dataset of the natural gas pipeline transportation network has been used in this paper, the result obtained was lower compared to the DDQN method.\\

\cite{kamble2019digitalization} & 2019 & Healthcare & -- & DoS Accuracy  & 98.5\%, using cloud database (Go-Daddy). It is able to fend against several types of network attacks, such as DoS and MiM. Only a limited number of network attacks and threats have been discussed.\\

\cite{otoum2021federated} & 2021 & Healthcare & SVM & Accuracy  & 98.5\% / 2.5\%, the recommended architecture would support both huge healthcare systems' intrusion detection as well as other wireless decentralized networks utilized in multiple real-world applications. Use only the earliest version of the dataset (CICIDS2017) to test the efficacy of the proposed model. \\

\cite{li2022online} & 2022 & Other & kNN & AUC & 82\%, used various datasets including their own dataset, in order to evaluate the efficiency. Then, the proposed approach was compared to five other ML algorithms, and the best result was obtained with kNN. Their proprietary dataset did not yield optimal results for the proposed model, and this model's capability is limited to detecting only DDoS attacks.\\

\cite{nie2021intrusion} & 2021 & Other & SRMF  & TPR  & 100\% / 17.24\%, the suggested approach accurately forecasts network traffic and detects intrusion, the best result obtained when it compared to sparse regularized matrix factorization
(SRMF). \\ 

\cite{zolotukhin2018artificial} & 2018 & Other & -- & --  & The approach was tested in SSH default password brute-forcing and slow HTTP DDoS attack, it demonstrated that the system is able to mitigate attacks in few seconds not only for detecting. \\
\hline
\end{tabular}
\end{table*}

\subsection{Other DRL-based IDS applications} 
The paper \cite{nie2021intrusion} describes a unique technique to intrusion detection in energy-efficient IoT devices. The researchers address the problem of detecting intrusions while taking into account the energy limits of IoT devices. They propose a DDPG based method for balancing intrusion detection accuracy and energy usage. Through continuous interactions with the environment, the algorithm learns an effective policy for IDS. The performance of the proposed technique demonstrates its effectiveness in detecting intrusions while utilizing the least amount of energy. The results indicate that the proposed DDPG-based algorithm has the ability to achieve effective and energy-efficient intrusion detection in green IoT systems through multiple factors like precision, TPR, FPR and F1-score. Moving on, the reference \cite{zolotukhin2018artificial} highlights the increasing number of malware targeting IoT devices and the limitations of traditional intrusion detection approaches in IoT networks. The proposed defense system utilizes SDN and network function virtualization (NFV) \cite{kheddar2022efficient} to enhance network security. The core component of the system is a DRL agent that evaluates the risks of potential attacks and takes optimal actions to mitigate them. Another reference \cite{li2022online} examines the challenge of intrusion detection in IoT systems and provides an online intrusion detection framework. To successfully identify intrusions in real-time, the system integrates full bayesian possibilistic clustering (FBPC) with ensembled fuzzy classifiers, to divide IoT data into clusters based on membership degrees. This allows the representation of data patterns to be flexible and granular. The data inside each cluster is then classified as normal or intrusive thanks to ensembled fuzzy classifiers.

\section{Future direction}
\label{sec4}
In order to fixe future research and development in the application of DRL for intrusion detection in IoT systems, several future possibilities are recommended in this work, including:
\begin{itemize}
    \item  Produce novel DRL models and strategies that can handle the particular challenges  of IoT intrusion detection such as privacy, security and adversarial attacks.

    \item Develop benchmark datasets specifically for DRL based intrusion detection in IoT systems.

    \item Combine DRL with other DL approaches, such as graph neural network (GNN), or similar approaches, in order to enhance the performance of IDSs in IoT environments.

    \item Examine the effectiveness of transfer learning methods in the context of DRL for intrusion detection in the IoT.

    \item Examine how DRL may be used to develop real-time, adaptive IDSs.

    \item Establish lightweight models and approaches that can detect and mitigate malicious activities and threats.
\end{itemize}

\label{sec6}

\section{Conclusion}
\label{sec5}
 In this research, a comprehensive survey on the application of DRL for intrusion detection in IoT domain was discussed. It covers a comprehensive examination of the existing research, illustrating the strengths and limitations of different strategies. The survey analyzes the possibility of DRL in boosting IDSs in IoT applications and highlights the main problems that need resolving. Overall, this paper underlines the opportunity of DRL to protect IoT systems against malicious threats, identifies future research targets, and advances knowledge of the existing SOTA techniques.
\label{sec7}

\section*{Acknowledgment}

The authors acknowledges that the study was partially funded by the PRFU-A25N01UN260120230001 grant from the Algerian Ministry of Higher Education and Scientific Research.

\balance
\bibliographystyle{IEEEtran}
\bibliography{references.bib}

\begin{thebibliography}{10}
\providecommand{\url}[1]{#1}
\csname url@samestyle\endcsname
\providecommand{\newblock}{\relax}
\providecommand{\bibinfo}[2]{#2}
\providecommand{\BIBentrySTDinterwordspacing}{\spaceskip=0pt\relax}
\providecommand{\BIBentryALTinterwordstretchfactor}{4}
\providecommand{\BIBentryALTinterwordspacing}{\spaceskip=\fontdimen2\font plus
\BIBentryALTinterwordstretchfactor\fontdimen3\font minus
  \fontdimen4\font\relax}
\providecommand{\BIBforeignlanguage}[2]{{%
\expandafter\ifx\csname l@#1\endcsname\relax
\typeout{** WARNING: IEEEtran.bst: No hyphenation pattern has been}%
\typeout{** loaded for the language `#1'. Using the pattern for}%
\typeout{** the default language instead.}%
\else
\language=\csname l@#1\endcsname
\fi
#2}}
\providecommand{\BIBdecl}{\relax}
\BIBdecl

\bibitem{kheddar2023deep}
H.~Kheddar, Y.~Himeur, and A.~I. Awad, ``Deep transfer learning for intrusion
  detection in industrial control networks: A comprehensive review,''
  \emph{Journal of Network and Computer Applications}, vol. 220, p. 103760,
  2023.

\bibitem{chang2022survey}
V.~Chang, L.~Golightly, P.~Modesti, Q.~A. Xu, L.~M.~T. Doan, K.~Hall, S.~Boddu,
  and A.~Kobusi{\'n}ska, ``A survey on intrusion detection systems for fog and
  cloud computing,'' \emph{Future Internet}, vol.~14, no.~3, p.~89, 2022.

\bibitem{umer2022machine}
M.~A. Umer, K.~N. Junejo, M.~T. Jilani, and A.~P. Mathur, ``Machine learning
  for intrusion detection in industrial control systems: Applications,
  challenges, and recommendations,'' \emph{International Journal of Critical
  Infrastructure Protection}, vol.~38, p. 100516, 2022.

\bibitem{rana2022intrusion}
P.~Rana, I.~Batra, A.~Malik, A.~L. Imoize, Y.~Kim, S.~K. Pani, N.~Goyal,
  A.~Kumar, and S.~Rho, ``Intrusion detection systems in cloud computing
  paradigm: Analysis and overview,'' \emph{Complexity}, vol. 2022, 2022.

\bibitem{vinolia2023machine}
A.~Vinolia, N.~Kanya, and V.~Rajavarman, ``Machine learning and deep learning
  based intrusion detection in cloud environment: A review,'' in \emph{2023 5th
  International Conference on Smart Systems and Inventive Technology
  (ICSSIT)}.\hskip 1em plus 0.5em minus 0.4em\relax IEEE, 2023, pp. 952--960.

\bibitem{liu2021review}
Q.~Liu, V.~Hagenmeyer, and H.~B. Keller, ``A review of rule learning-based
  intrusion detection systems and their prospects in smart grids,'' \emph{IEEE
  Access}, vol.~9, pp. 57\,542--57\,564, 2021.

\bibitem{kheddarASR2023}
H.~Kheddar, Y.~Himeur, S.~Al-Maadeed, A.~Amira, and F.~Bensaali, ``Deep
  transfer learning for automatic speech recognition: Towards better
  generalization,'' \emph{Knowledge-Based Systems}, vol. 277, p. 110851, 2023.

\bibitem{kheddar2023deepSteg}
H.~Kheddar, M.~Hemis, Y.~Himeur, D.~Meg{\'\i}as, and A.~Amira, ``Deep learning
  for steganalysis of diverse data types: A review of methods, taxonomy,
  challenges and future directions,'' \emph{Neurocomputing}, p. 127528, 2024.

\bibitem{wang2022policy}
Y.~Wang and S.~Zou, ``Policy gradient method for robust reinforcement
  learning,'' in \emph{International Conference on Machine Learning}.\hskip 1em
  plus 0.5em minus 0.4em\relax PMLR, 2022, pp. 23\,484--23\,526.

\bibitem{benaddi2020deep}
H.~Benaddi, K.~Ibrahimi, A.~Benslimane, and J.~Qadir, ``A deep reinforcement
  learning based intrusion detection system (drl-ids) for securing wireless
  sensor networks and internet of things,'' in \emph{Wireless Internet: 12th
  EAI International Conference, WiCON 2019, TaiChung, Taiwan, November 26--27,
  2019, Proceedings 12}.\hskip 1em plus 0.5em minus 0.4em\relax Springer, 2020,
  pp. 73--87.

\bibitem{tu2022intrusion}
J.~Tu, W.~Ogola, D.~Xu, and W.~Xie, ``Intrusion detection based on generative
  adversarial network of reinforcement learning strategy for wireless sensor
  networks,'' \emph{International Journal of Circuits, Systems and Signal
  Processing}, vol.~16, pp. 478--482, 2022.

\bibitem{diddigi2018novel}
R.~B. Diddigi, K.~Prabuchandran, and S.~Bhatnagar, ``Novel sensor scheduling
  scheme for intruder tracking in energy efficient sensor networks,''
  \emph{IEEE Wireless Communications Letters}, vol.~7, no.~5, pp. 712--715,
  2018.

\bibitem{madhurireinforcement}
K.~S. Madhuri and J.~Mungara, ``Reinforcement learning for intrusion detection
  and improving optimal route by cuckoo search in wsn.''

\bibitem{ramana2022ambient}
T.~Ramana, M.~Thirunavukkarasan, A.~S. Mohammed, G.~G. Devarajan, and S.~M.
  Nagarajan, ``Ambient intelligence approach: Internet of things based decision
  performance analysis for intrusion detection,'' \emph{Computer
  Communications}, vol. 195, pp. 315--322, 2022.

\bibitem{nguyen2021federated}
T.~G. Nguyen, T.~V. Phan, D.~T. Hoang, T.~N. Nguyen, and C.~So-In, ``Federated
  deep reinforcement learning for traffic monitoring in sdn-based iot
  networks,'' \emph{IEEE Transactions on Cognitive Communications and
  Networking}, vol.~7, no.~4, pp. 1048--1065, 2021.

\bibitem{kalnoor2021markov}
G.~Kalnoor \emph{et~al.}, ``Markov decision process based model for performance
  analysis an intrusion detection system in iot networks,'' \emph{Journal of
  Telecommunications and Information Technology}, 2021.

\bibitem{otoum2021ids}
Y.~Otoum and A.~Nayak, ``As-ids: Anomaly and signature based ids for the
  internet of things,'' \emph{Journal of Network and Systems Management},
  vol.~29, pp. 1--26, 2021.

\bibitem{tharewal2022intrusion}
S.~Tharewal, M.~W. Ashfaque, S.~S. Banu, P.~Uma, S.~M. Hassen, and M.~Shabaz,
  ``Intrusion detection system for industrial internet of things based on deep
  reinforcement learning,'' \emph{Wireless Communications and Mobile
  Computing}, vol. 2022, pp. 1--8, 2022.

\bibitem{kamble2019digitalization}
P.~Kamble and A.~Gawade, ``Digitalization of healthcare with iot and
  cryptographic encryption against dos attacks,'' in \emph{2019 international
  conference on contemporary computing and informatics (IC3I)}.\hskip 1em plus
  0.5em minus 0.4em\relax IEEE, 2019, pp. 69--73.

\bibitem{otoum2021federated}
S.~Otoum, N.~Guizani, and H.~Mouftah, ``Federated reinforcement
  learning-supported ids for iot-steered healthcare systems,'' in \emph{ICC
  2021-IEEE International Conference on Communications}.\hskip 1em plus 0.5em
  minus 0.4em\relax IEEE, 2021, pp. 1--6.

\bibitem{li2022online}
F.-Q. Li, R.-J. Zhao, S.-L. Wang, L.-B. Chen, A.~W.-C. Liew, and W.~Ding,
  ``Online intrusion detection for internet of things systems with full
  bayesian possibilistic clustering and ensembled fuzzy classifiers,''
  \emph{IEEE Transactions on Fuzzy Systems}, vol.~30, no.~11, pp. 4605--4617,
  2022.

\bibitem{nie2021intrusion}
L.~Nie, W.~Sun, S.~Wang, Z.~Ning, J.~J. Rodrigues, Y.~Wu, and S.~Li,
  ``Intrusion detection in green internet of things: a deep deterministic
  policy gradient-based algorithm,'' \emph{IEEE Transactions on Green
  Communications and Networking}, vol.~5, no.~2, pp. 778--788, 2021.

\bibitem{zolotukhin2018artificial}
M.~Zolotukhin and T.~H{\"a}m{\"a}l{\"a}inen, ``On artificial intelligent
  malware tolerant networking for iot,'' in \emph{2018 IEEE Conference on
  Network Function Virtualization and Software Defined Networks
  (NFV-SDN)}.\hskip 1em plus 0.5em minus 0.4em\relax IEEE, 2018, pp. 1--6.

\bibitem{kheddar2022efficient}
H.~Kheddar, Y.~Himeur, S.~Atalla, and W.~Mansoor, ``An efficient model for
  horizontal slicing in 5g network using practical simulations,'' in \emph{2022
  5th International Conference on Signal Processing and Information Security
  (ICSPIS)}.\hskip 1em plus 0.5em minus 0.4em\relax IEEE, 2022, pp. 158--163.

\end{thebibliography}

\end{document}